\documentclass[journal]{IEEEtran}

\usepackage{comment}
\usepackage{mathtools}
\usepackage{amsmath}
\usepackage{graphicx}
\usepackage{epstopdf}

\usepackage{hyperref} 
\usepackage{cite}
\usepackage{amsmath}
\usepackage{color}

\usepackage{graphicx}
\usepackage{mathptmx}      
\usepackage{latexsym}
\usepackage[ruled,vlined]{algorithm2e}

\usepackage[misc]{ifsym}

\DeclareMathOperator{\atantwo}{atan2}

\begin{document}

\title{Learning Sparse Spatial Codes for Cognitive Mapping Inspired by Entorhinal-Hippocampal Neurocircuit}

\author{Taiping Zeng, XiaoLi Li, and Bailu Si
\thanks{T. Zeng is with Institute of Science and Technology for Brain-Inspired  Intelligence, Fudan University, Shanghai, China and Key Laboratory of Computational Neuroscience and Brain-Inspired Intelligence (Fudan University), Ministry of Education, China (e-mail:zengtaiping.ac@gmail.com).}
\thanks{X. Li, State Key Laboratory of Cognitive Neuroscience and Learning and IDG/McGovern Institute for Brain Research, Beijing Normal University, Beijing, China (e-mail:xiaoli@bnu.edu.cn).}
\thanks{B. Si is with School of Systems Science, Beijing Normal University, 100875, China (e-mail:bailusi@bnu.edu.cn).}
\thanks{Correspondence should be addressed to Bailu Si (bailusi@bnu.edu.cn).}
}


\maketitle

\begin{abstract}
The entorhinal-hippocampal circuit plays a critical role in higher brain functions, especially spatial cognition. Grid cells in the medial entorhinal cortex (MEC) periodically fire with different grid spacing and orientation, which makes a contribution that place cells in the hippocampus can uniquely encode locations in an environment. 
But how sparse firing granule cells in the dentate gyrus are formed from grid cells in the MEC remains to be determined.  
Recently, the fruit fly olfactory circuit provides a variant algorithm (called locality-sensitive hashing) to solve this problem.
To investigate how the sparse place firing generates in the dentate gyrus can help animals to break the perception ambiguity during environment exploration, we build a biologically relevant, computational model from grid cells to place cells. The weight from grid cells to dentate gyrus granule cells is learned by competitive Hebbian learning. 
We resorted to the robot system for demonstrating our cognitive mapping model on the KITTI odometry benchmark dataset.  
The experimental results show that our model is able to stably, robustly build a coherent semi-metric topological map in the large-scale outdoor environment.
The experimental results suggest that the entorhinal-hippocampal circuit as a variant locality-sensitive hashing algorithm is capable of generating sparse encoding for easily distinguishing different locations in the environment. 
Our experiments also provide theoretical supports that this analogous hashing algorithm may be a general principle of computation in different brain regions and species. 
\end{abstract}


\section{Introduction}
\label{intro}
Spatial cognition endows many animals with impressive navigation capabilities, who can travel long distances to find mates, shelter, food, and water, and then correctly return to their home in the large-scale environment.
Tolman proposed that a mental map-like representation in the brain, called cognitive map, allows animals to learn spatial information and performs space-dependent cognitive tasks, such as exploration, mapping, localization, planning paths, and navigating paths~\cite{tolman_cognitive_1948}. 
The discovery of place cells with striking spatial selectivity in the hippocampus provides evidence for the hypothesis of cognitive map~\cite{okeefe_hippocampus_1971, okeefe_hippocampal_1978}. 
Head direction cells (HD cells) were originally discovered in the rodent dorsal presubiculum, which fire maximally when an animal's head faces a particular direction~\cite{taube_head-direction_1990}. This preferred direction is independent of the animal's position.
Grid cells in the medial entorhinal cortex (MEC) fired in multiple discrete locations in the environment, which expressed a hexagon grid pattern across the whole explored environment~\cite{hafting_microstructure_2005}. The scale of the grid pattern increased from dorsal to ventral in the MEC, which may provide representations for different scale environments. 
It was clear that place cells in the hippocampus and grid cells in the entorhinal cortex were an essential part of neural circuits in spatial representation. 

The hippocampus is considered to be an essential part of neural circuits for spatial representation and navigation~\cite{okeefe_hippocampal_1978}. The dentate gyrus, as the input region of the hippocampus, preprocesses incoming information from the superficial layers of the entorhinal cortex, and outputs to the CA3 region of the hippocampus. Cells with multiple spatial fields are converged into cells with single specific spatial fields. Additionally, the dentate gyrus also achieves pattern separation functions. Similar incoming patterns are transformed into completely different output patterns. Finally, sparse firing granule cells in the dentate gyrus are thought to undertake pattern separations. At any one time, only a very small part of granule cells are firing~\cite{jonas_structure_2014}. 

But, unfortunately, how place cells are formed from diverse inputs remains to be determined~\cite{moser2015place_csh}. Numerous computational models have been proposed to account for this mystery.
Neurophysiological experiments and anatomical connectivity imply that grid cells in the MEC are likely to provide principle inputs to place cells in the hippocampus. Grid cells in the superficial layers of the entorhinal cortex is the main cortical input to the hippocampus. Many researchers proposed that place fields are generated by a linear combination of periodic grid firing fields with different grid spacing and orientation~\cite{okeefe_dual_2005, solstad_grid_2006, mcnaughton_path_2006}. 
Considering individual grid patterns with different scale as different Fourier components, the place field of the current place cell is formed by enhancing the central peak and canceling others~\cite{ormond_place_2015, kubie_spatial_2015}. Nevertheless, grid cells can converge their neural activity into a single peak only when grid phases of all input grid cells are aligned on the central peak, and the extra peaks can be suppressed. 
Meanwhile, Rolls et al.~\cite{rolls2006entorhinal} proposed that place responses may be generated from a process of self-organized plasticity. The connections from grid cells in the MEC to granule cells in the dentate gyrus are learned through competitive Hebbian learning~\cite{si2009role, monaco2011modular}.
The dentate gyrus granule cells are turned into place cells with a single peak, which have multiple small firing fields with the initial random connection weights.

Although a lot of computational models describing the interaction between the entorhinal cortex and hippocampus have been proposed, how these models response to the real physical environment in robotic experiment needs to be further explored.  
A model from grid cells and visual place cells to multimodal place cells was proposed to reproduce neurobiological experimental results. Although the model was tested in robotic experiments, it seldom helps improve the performance of the mobile robot~\cite{jauffret2015grid}. Competitive Hebbian learning was applied to compute the place cell response from grid cell population activities based on existed continuous attractor network~\cite{yoram_burak_accurate_2009} for building cognitive map~\cite{yuan_entorhinal-hippocampal_2015,hu_hebbian_2016}. The robotic system is tested in small-scale indoor environments.
Population activity of place cells is generated from grid patterns with different scale. 
And yet, the modeled place cells do not identify specific regions in the hippocampus. The place cell representation does also not use sparse coding scheme. An amount of grid cell layers are applied to form place cell activity, which is quite opposite with neurobiological experimental facts.
 
Actually, the mechanism that the dentate gyrus produces sparse firing patterns to separate incoming periodic grid spatial patterns still remains unclear and requires to be further investigated in the real robotic experiments. Recently, a neural circuit processing odors in the fruit fly olfactory system is thought to provide a variant algorithm (called locality-sensitive hashing, LSH)~\cite{datar2004locality} to solve the pattern separation problem~\cite{dasgupta_neural_2017}. A 'tag' for each odor is generated by the fly olfactory circuit. The tags are sparsely represented by a small fraction of the neurons for each odor. Two randomly selected, dissimilar odors share few active neurons in the neural circuit, which is beneficial to easily distinguish different odors. This fly algorithm strategy may also appear in mouse olfactory circuit and rat cerebellum. Projection from the entorhinal cortex to dentate gyrus in the rat hippocampus is poorly known~\cite{dasgupta_neural_2017}.

In this work, following the discovery of the fruit fly olfactory algorithm~\cite{dasgupta_neural_2017}, we developed a cognitive mapping model for mobile robots, according to anatomical structures and known hypotheses of the entorhinal-hippocampal circuit. Our model integrates local view cells, HD cells, conjunction grid cells in the MEC layer III, V, and VI~\cite{Sargolini2006ConjunctiveRO}, grid cells in the MEC layer II, and place cells in the dentate gyrus together for building cognitive maps. Both HD cells and conjunctive grid cells are modeled by continuous attractor networks with the same mechanisms. HD cell model conjunctively represents the rotational velocity and head directions of the animals. In a similar way, grid cell model conjunctively encodes scaled translational velocities and possible period positions in the two-dimensional environment. More details about HD cell and conjunctive grid cell models can be found in our previous work~\cite{zeng_cognitive_2017}. Three conjunctive grid cell layers mimic the entorhinal cortex from dorsal to ventral with scaled speed inputs from large gain to small gain~\cite{mcnaughton_path_2006}. Conjunctive grid cell activities in the deep MEC layers are summed along velocity dimensions to generate grid cell activities in the superficial MEC layers. The weight from grid cells to dentate gyrus is randomly initialized and learned through competitive Hebbian learning. After a variant algorithm in the fruit fly olfactory circuit, the input (grid cells) dimensionality is expanded after projection into the output (dentate gyrus granule cells)~\cite{dasgupta_neural_2017}. 
\begin{figure*}[!ht]
\centering
\includegraphics[width=6.8in]{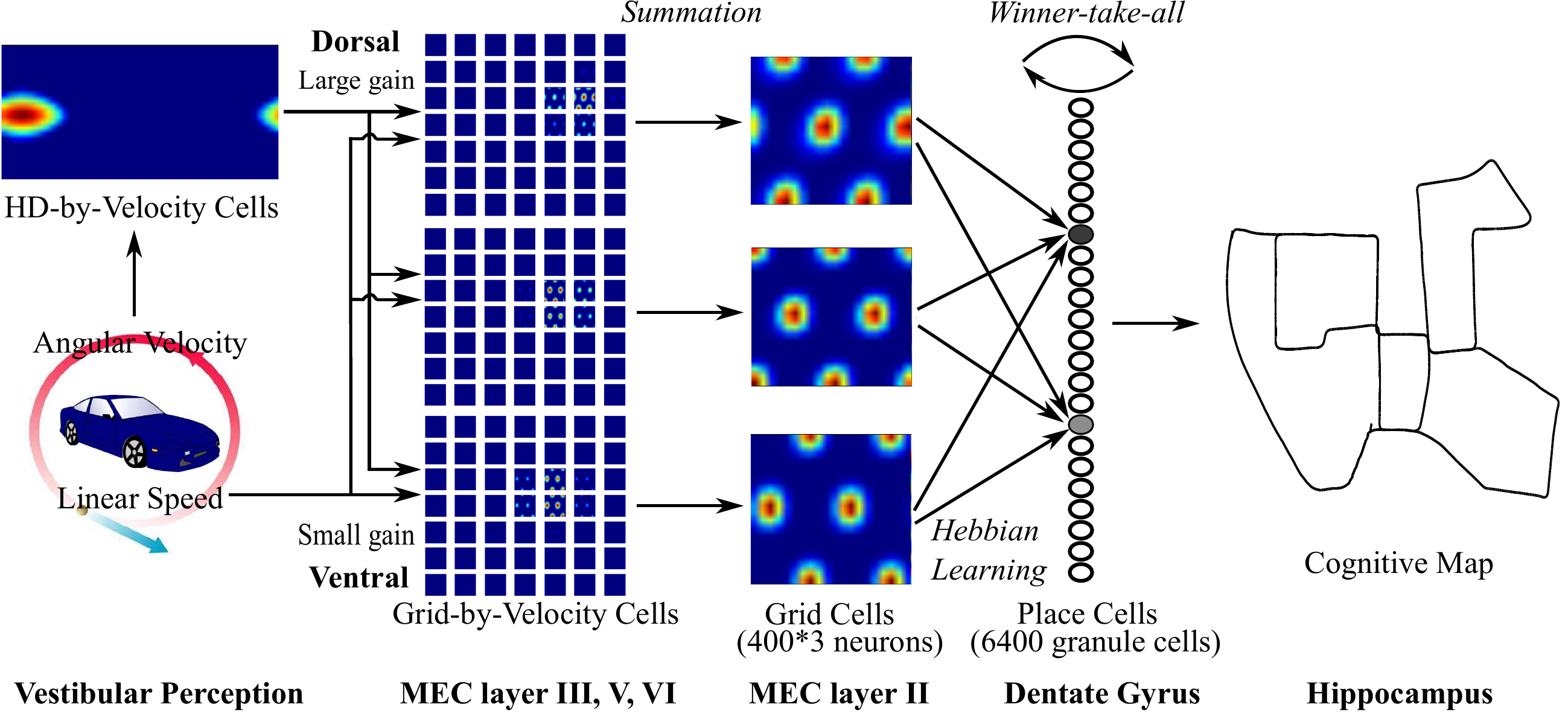}
\caption{The neural network architecture of the model and the diagram of information flow. The HD-by-velocity cells, updated by angular velocity, represent head directions. Three grid-by-velocity cells receiving translational velocity with different gains, converted from linear speed and the head direction representations, provide positional representation. The activity of grid cells is generated from corresponding grid-by-velocity cells by summation along velocity dimensions. The sparse place firing responses are generated from three grid cells in the dentate gyrus granule cells by the Winner-take-all mechanism with competitive Hebbian learning rules and controlling the sparsity of dentate gyrus activity. The neural encodes are read from the spatial memory network to build a cognitive map.}
\label{fig:architecture}
\end{figure*}
Winner-take-all (WTA) is adopted to make a very sparse representation by controlling the sparsity of the dentate gyrus granule cells. The place firing fields are generated by the dentate gyrus granule cells with a variant algorithm, which is analogous with the fruit fly olfactory circuit (a variant locality-sensitive hashing)~\cite{dasgupta_neural_2017}. The sparse firing neural activity of place cells, namely dentate gyrus granule cells, can uniquely encode the robot positions in the explored environment. We demonstrate our cognitive mapping model on the KITTI odometry benchmark dataset~\cite{Geiger2012CVPR}, which is recorded from a car with relatively high speed in urban and highway environments.
The experimental results show that our model is able to stably, robustly build a coherent semi-metric topological map in the large-scale outdoor environment. 

This paper makes the following three specific contributions.
First, based on the anatomical structures and the cognitive functions of entorhinal-hippocampal circuits of mammalian brains, we proposed a model as a neural substrate to validate the neural activity of entorhinal-hippocampal circuits, which organically integrates HD cells, conjunction grid cells in the MEC layer III, V, and VI, grid cells in the MEC layer II, and place cells in the dentate gyrus together. The sparse place firing responses are generated in the dentate gyrus granule cells by the Winner-take-all mechanism with competitive Hebbian learning rules and controlling the sparsity of dentate gyrus activity.
Second, we implemented our entorhinal-hippocampal model on the robotic system, and demonstrated its performance on the KITTI odometry benchmark dataset. Moreover, we expand our cognitive mapping system to process stereo visual inputs with a proved direct sparse visual method. Our system successfully generates a coherent semi-metric topological map in the large-scale outdoor environment only from a moving stereo camera.  
Third, we further proved the hypothesis that this analogous algorithm (i.e., locality-sensitive hashing) from grid cells in the entorhinal cortex to dentate gyrus granule cells in the hippocampus may be a general principle of computation in the brain with different regions and species. 

\section{Method}
\label{Method}
In this work, we proposed a model for exploring spatial cognition that biologically mimics the cognitive function and anatomical structure of entorhinal-hippocampal circuits of mammalian brains. 
Vestibular perception is implemented by a proved direct sparse visual odometry method to provide linear speed and angular velocity~\cite{wang_stereo_2017}. Path integration is performed by the HD cells and the conjunctive grid cells in the MEC. Pattern separation is realized by the projection from three conjunctive grid cell layers with different speed gains to dentate gyrus granule cells to generate place firing fields. The local view cells in the model correspond to the retrosplenial cortex or visual cortex and provide visual information to the entorhinal-hippocampal circuit. Cognitive map corresponds to the hippocampus to describe an internal spatial representation among salient landmarks in an environment. In the following, the core components of the proposed cognitive mapping model is described for showing high mapping performance in the large-scale environment.

\subsection{Neural Network Architecture}
The architecture of the proposed model is shown in Figure~\ref{fig:architecture}. Angular velocity and linear speed are estimated by a traditional direct sparse visual odometry, which not includes biological inspiration. The proved odometry achieves good performance both in terms of accuracy and robustness from a single moving stereo camera~\cite{wang_stereo_2017}.  

A ring attractor in the neural space is formed by the HD cells through population coding. The head direction of the robot in the physical environment is presented by the phase of an activity bump.  
When a real angular velocity value is fed into the HD cell network from the real robot, a subset of HD cells would be activated. Then the activity bump would rotate intrinsically with the same input angular velocity through attractor dynamics. 
Confined by the periodic boundary conditions, the conjunctive grid cells generates a torus attractor in the neural space. The HD cells and linear speed are tuned to activated a subset of conjunctive grid cells. The grid activity bump moves through intrinsic attractor dynamics with a velocity proportional to the movement of the animal in the physical environment. Conjunctive grid cells simultaneously include the properties of grid cells and speed cells~\cite{kropff_speed_2015}, and represent the scaled velocity and the possible periodic position. 

Here, three conjunctive grid cell layers with different speed gains from large to small generate three different grid patterns, corresponding to neurons in the MEC from dorsal to ventral~\cite{mcnaughton_path_2006}. Three corresponding grid cell layers in the MEC layer II are shaped by removing the velocity dimensions of the three conjunctive grid cell layers. Every grid layer has 400 grid cells. The place firing responses in the dentate gyrus granule cells are generated from the three grid layers. First, the weight from three grid cell layers to dentate gyrus granule cells are randomly initialized. Then, the weight is learned by competitive Hebbian learning to form place fields. During the learning process, winner-take-all is adopted to control the sparsity of dentate gyrus granule cells. Thus, very sparse place cell codings emerge in the dentate gyrus to separate similar grid patterns for distinguishing positions of the robot in the physical environment.

\subsection{HD-by-Velocity Cell Model}

Angular velocity is integrated by HD-by-velocity cell network representing a one-dimensional head direction. Head direction and rotation are conjunctively encoded in the HD-by-velocity cell network.

\subsubsection{Neural Representation of Head Direction and Angular Velocity}

Each unit in the network is labeled by its coordinate ($\theta , \nu$) on a two dimensional neural manifold. $\theta \in [0, 2\pi)$ is the internal representation of head directions with periodic boundary condition in the environment. Angular velocity is encoded by $\nu \in [-L_r, L_r]$. Both $\nu$ and $\theta$ are dimensionless quantities in the neural space, which are related to the real angular velocity and the head direction of the robot. The connection weights are designed between units to generate a single stable bump of activity both in the direction dimension $\theta$ and in the rotation dimension $\nu$.
The strength of the connection between a presynaptic unit $(\theta',\nu')$ and a postsynaptic unit $(\theta,\nu)$ can be defined as
\begin{equation}
J(\theta,\nu|\theta',\nu')=J_0+J_1\cos\left(\theta - \theta' - \nu'\right)\cos\left(\lambda (\nu - \nu')\right).
\end{equation}
where $J_0<0$ is a uniform inhibition, $J_1>0$ describes the interaction strength, and $\lambda$ defines the spread of velocity tuning. 
As the center of the postsynaptic unit is not at $\theta'$ but at $\theta'+\nu'$, the connection weight from the unit at $\theta'$ to the unit at $\theta$ in the direction dimension is asymmetric. The asymmetric weights cause the bump to move along the direction dimension with a velocity determined by $\nu'$.

\subsubsection{Network Dynamics}
The bump activity of the network is driven by by velocity and sensory inputs. The firing rate $m(\theta,\nu)$ of the unit at coordinate ($\theta , \nu$) can be defined as 
\begin{equation}
\begin{split}
\displaystyle
\tau \dot{m}(\theta,\nu) &= -m(\theta,\nu) \\
&+ f\left(\iint D\theta D\nu J(\theta,\nu|\theta',\nu')m(\theta',\nu') + I_{\nu} + I_{view}\right)\label{eq:HD_dynamics},
\end{split}
\end{equation}
where $I_{\nu}$ and $I_{view}$ are the velocity tuned input and the calibration current injected by local view cells respectively, which we will explain in detail below. $\tau$ is the time constant set to 10 ms. $f(x)$ is a threshold-linear function: $f(x) \equiv [x]_+ = x$ when $x>0$ and 0 otherwise. The shorthand notations are used $\displaystyle \int D\theta = \frac{1}{2\pi}\int_{0}^{2\pi} d\theta$, and $\displaystyle \int D\nu = \frac{1}{2L_r}\int_{-L_r}^{L_r} d\nu$.

\subsubsection{Angular Velocity Inputs}\label{sec:v-input-hd}
To integrate the real angular velocity of the robot, it must be mapped onto the neural manifold of the HD network at the appropriate position. 
Given an external angular velocity $V$, the desired bump location in the $\nu$ dimension can be written as \cite{si_continuous_2014}
\begin{equation}
u(V)=\arctan\left(\tau V \right). \label{eq:angular-velocity-map}
\end{equation}
Here $\tau$ is the time constant defined in Equation~\ref{eq:HD_dynamics}. 
The velocity input to the HD-by-velocity units is simply modeled by a tuning function of Gaussian shape
\begin{equation}
I_{\nu}(\nu|V)=I_r\left[1-\epsilon_r + \epsilon_r,\exp\left(-\frac{(\nu-u(V))^2}{2\sigma_r^2}\right)\right]. \label{eq:angular-v-input}
\end{equation}
Here $I_r$ is the amplitude of the rotational velocity input, $\epsilon_r$ defines the strength of velocity tuning, and $\sigma_r$ is the sharpness of the velocity tuning.

\subsubsection{Estimation of Head Direction}
The head direction and angular velocity of the robot in the physical environment are encoded by the activity bump on the $\theta$ axis and $\nu$ axis. Fourier transformations are utilized to recover the head direction and angular velocity of the robot from the neural activity of the network
\begin{equation}
\psi = \angle(\displaystyle \iint m(\theta, \nu) \exp(i \theta) D\theta D\nu),\label{eq:hd-estimation}
\end{equation}
\begin{equation}
\phi = \frac{\displaystyle \angle(\iint m(\theta, \nu) \exp(i \lambda \nu) D\theta D\nu)}{\lambda}.\label{eq:HD_speed_est}
\end{equation}
Here $i$ is the imaginary unit, and function $\angle (Z)$ takes the angle of a complex number $Z$. $\psi \in [0, 2\pi)$ is the estimated phase of the bump in the direction axis of the neural space, corresponding to the head direction of the robot in the physical environment. $\phi \in (-L_r,L_r)$ is the estimated phase of the bump in the velocity axis of the neural space, and can be recovered to the angular velocity of the robot in the physical space by inverting Equation~\ref{eq:angular-velocity-map} 
\begin{equation}
V = \frac{\tan(\phi)}{\tau}.
\end{equation}
Note that $L_r$ should be selected large enough, so that the recovered velocity $V$ is capable of representing all possible angular velocities of the robot.

\subsection{Grid-by-Velocity Cell Model}
Then, our HD-by-velocity cell model is expanded to do path integration in the two-dimensional environment. Two-dimensional spatial locations and two-dimensional velocity are represented in the grid-by-velocity cell network. The units in the network are wired with appropriate connection profiles, so that the hexagonal grid firing pattern is created and translated in the spatial dimension of the neural manifold.

\subsubsection{Neural Representation of Position and Velocity}
Units in the grid-by-velocity network is labeled by coordinates $(\vec{\theta},\vec{\nu})$ in a four dimensional neural space.  $\vec{\theta} = (\theta_x,\theta_y)$ represents two dimensional positions with periodic boundary conditions in the environment, i.e. $\theta_x,\theta_y \in [0,2\pi)$. $\nu_x$ and $\nu_y$ are chosen in $[-L_t,L_t]$. $\vec{\nu} = (\nu_x,\nu_y)$ encodes the velocity components in the environment. The connection weights from unit $(\vec{\theta'},\vec{\nu'})$ to $(\vec{\theta},\vec{\nu})$ is described as 
\begin{equation}
\begin{split}
\displaystyle
J(\vec{\theta},\vec{\nu}|\vec{\theta'},\vec{\nu'}) &=J_0 + J_k\cos\left(k\sqrt{\sum \limits_{j\in\{x,y\}}||\theta_j -\theta'_j - \nu'_j ||^2 } \right) \\
&\cos\left(\lambda\sqrt{\sum \limits_{j\in\{x,y\}} (\nu_j - \nu'_j)^2}\right)\label{eq:grid_weight},
\end{split}
\end{equation}
where integer $k = 2$, is chosen so that the network accommodates two bumps both in $\theta_x$ axis and in $\theta_y$ axis. There is only one bump in each of the velocity dimensions however. $||d||$ is the distance on a circle: $||d|| = \textrm{mod}(d + \pi, 2\pi) - \pi$, and $\textrm{mod}(x,y) \in [0,y)$ gives x modulo y. 

\subsubsection{Network Dynamics}
Although the manifold structure of the grid-by-velocity cell network are different with the HD-by-velocity cell network, they share the same intrinsic dynamics as in Equation~\ref{eq:HD_dynamics}
\begin{equation}
\begin{split}
\displaystyle
\tau \dot{m}(\vec{\theta},\vec{\nu}) &= -m(\vec{\theta},\vec{\nu}) \\
&+ f\left(\iint D\vec{\theta} D\vec{\nu}J(\vec{\theta},\vec{\nu}|\vec{\theta'},\vec{\nu'})m(\vec{\theta'},\vec{\nu'}) + I_{\nu} + I_{view}\right).
\end{split}
\end{equation}
Note that $\displaystyle \int D\vec{\theta} = \frac{1}{4\pi^2}\int_{0}^{2\pi}\int_{0}^{2\pi} d\theta_x d\theta_y$, and \newline
 $\displaystyle \int D\vec{\nu} = \frac{1}{4L_t^2}\int_{-L_t}^{L_t}\int_{-L_t}^{L_t} d\nu_x d\nu_y$.

\subsubsection{Translational Velocity Inputs}\label{sec:v-input-grid}
For performing accurate path integration, the input velocity of the robot in the physical environment should be proportional to the velocity of the moving bumps in the neural manifold. The activity bumps are pinned on appropriate positions on the velocity axes according to the tuned velocity inputs, so that the bumps move with the desired velocity.

The translational velocity $\vec{V} = (V_x,V_y)$ of the robot is calculated from the head direction estimated from HD-by-velocity units (Equation~\ref{eq:hd-estimation}) and linear speed by projecting to the axes of the reference frame.
The running speed is encoded by the speed cells in the MEC of the rodent brain. Given the translational velocity of the robot, the desired positions on the velocity axes in the neural space are given by $\vec{u}(\vec{V})$\cite{si_continuous_2014}
\begin{equation}
\vec{u}(\vec{V})=\frac{1}{k}\arctan\left(\frac{2\pi \tau \vec{V}}{S}\right), \label{eq:grid_velocity_input}
\end{equation}
where the function $\arctan$ operates on each dimension of $\vec{V}$. S is a scaling factor between the external velocity of the robot in the physical environment and the velocity of the bumps in the neural space. S determines the spacing between the fields of grid firing pattern in the environment.

The velocity-tuned inputs to the grid-by-velocity units are tuned by a Gaussian form for simplicity
\begin{equation}
I_\nu(\vec{\nu}|\vec{V})=I_t\left[1-\epsilon + \epsilon \,\exp\left(-\frac{|\vec{\nu}-\vec{u}(\vec{V})|^2}{2\sigma_t^2}\right)\right],\label{eq:grid_velocity_tunning}
\end{equation}
where $|\cdot|$ is the Euclidean norm of a vector. $I_t$ is the amplitude of the translational velocity input.    

\subsection{Model from Grid Cells to Dentate Gyrus Granule Cells}
Three grid layers in the MEC layer II are projected to form sparse place firing patterns in the dentate gyrus. Sparse place representation is able to realize pattern separation between similar grid patterns. It is similar to a variant LSH. 

\subsubsection{Model}
Grid activities $m(\vec{\theta})$ are summed from conjunctive grid activities $m(\vec{\theta},\vec{\nu})$ along velocity axes $\vec{\nu}$. The total number of all three grid layers is N = 1,200 units. The total number of place cells in the dentate gyrus is M = 6,400 units. Each place unit receives input from all grid units in three grid layers. The firing rate $p_{i}$ of a place unit can be defined as
\begin{equation}
\begin{split}
\displaystyle
p_{i} = f\left(\sum_{j = 1}^{M} w_{ij} m_{j}(\vec{\theta}) - \delta \right),
\end{split}
\end{equation}
where $j$ is the index of grid units, $w_{ij}$ is synaptic weight from grid unit $j$ to place unit $i$; $f(x)$ is a threshold-linear function, $f(x) = x$ when $x > 0$ and $0$ otherwise. $\delta$ is the threshold of the place cell network, calculated each time to ensure that the mean and sparsity of the place unit activity are both equal to a preset constant value $a$
\begin{equation}
\begin{split}
\displaystyle
\frac{\sum_{i=1}^M p_i}{M} = a,
\end{split}
\end{equation}
\begin{equation}
\begin{split}
\displaystyle
\frac{(\sum_{i=1}^M p_i/M)^2}{\sum_{i=1}^M p^2_i/M} = a.
\end{split}
\end{equation}

\subsubsection{Hebbian Learning Rules}
The weights are learned through a competitive Hebbian learning rule, which can be described as
\begin{equation}
\begin{split}
\displaystyle
w_{ij}(t+1) = f\left(w_{ij}(t) + \epsilon p_{i} \left(m_{j}(\vec{\theta}) - \langle m_{j}(\vec{\theta} \right) \rangle \right),
\end{split}
\end{equation}
where $t$ denotes the current time step, $t+1$ is the next time step; $f(x)$ is also a threshold-linear function, $f(x) = x$ when $x > 0$ and $0$ otherwise; $\langle \cdot \rangle$ represents the average of the elements. $\epsilon$ is a positive learning rate, which is set to $0.000001$.
The weights are randomly initialized in the range $[0,1]$ before first learning. At each time step, the weights are normalized by 
\begin{equation}
\begin{split}
\displaystyle
\sqrt{\sum_j w^2_{ij}(t)} = 1.
\end{split}
\end{equation}

Combination of the threshold of the place cell network $\delta$ and Hebbian learning rules, only a certain amount of place units are activated represented by the WTA mechanism.

\subsection{Direct Sparse Stereo Visual Odometry}

\begin{figure}[!ht]
\centering
\includegraphics[width=3.4in]{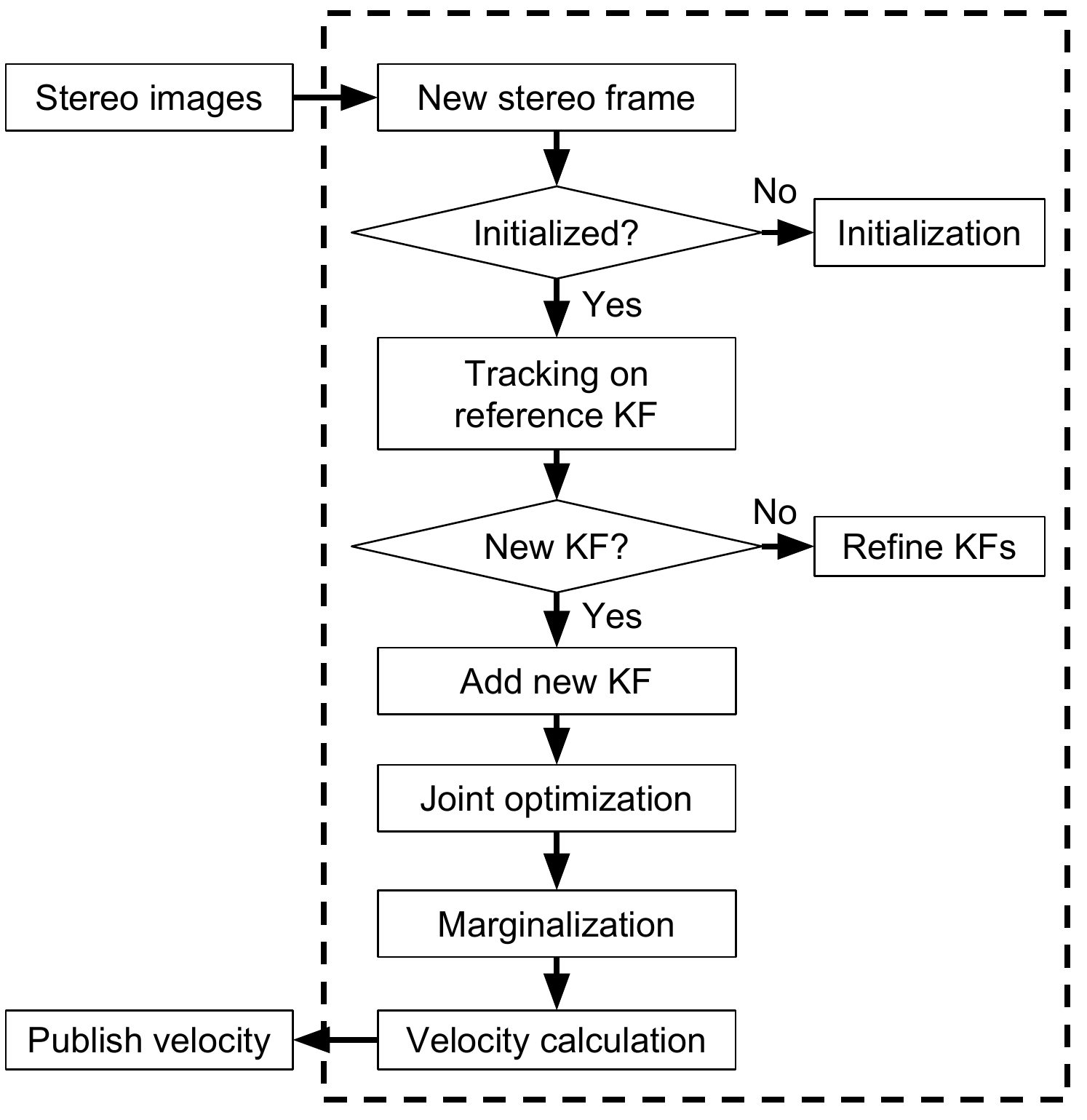}
\caption{The diagram of the direct sparse stereo visual odometry.}
\label{fig:vo_sch}
\end{figure}

In order to obtain an undistorted cognitive map, the accurate estimation of velocity is urgently required. Previous researches~\cite{ball_openratslam_2013,zeng_cognitive_2017} are only capable of performing rough velocity estimation with the scanline intensity profiles. In our SLAM system, we adopted a proved visual odometry with more accuracy and robustness~\cite{engel_direct_2017, wang_stereo_2017}. In the next, we would only provide an overview of the visual odometry.

The visual odometry receives the stereo images and publishes current velocity information to other nodes by ROS message types. The direct sparse stereo visual odometry processes stereo images shown in Fig~\ref{fig:vo_sch}. First, depth estimation from static stereo matching is employed to initialize the system. Second, every new frame is tracked with respect to their reference keyframe (KF). Third, Whether the new keyframe is added to the current active window is determined by scene or illumination changes.  Fourth, a joint optimization is performed for all keyframes in the current active window, including keyframes' poses, affine brightness parameters, the depths of all the observed selected points, and camera intrinsics. Fifth, to prevent the growth of the size in the active window, active points not observed by the two latest keyframes and hosted in the old keyframe, as well as the old keyframe are marginalized out. Finally, the current velocity is estimated and published to other nodes.

In this section, the same denotation applied from~\cite{engel_direct_2017}. Light, bold lower-case letters and bold upper-case letters are utilized to denote scalars ($u$), vectors (\textbf{t}) and matrices (\textbf{R}) respectively. 
Functions ($I$) is represented by Light upper-case letters. Camera calibration matrices are defined by \textbf{K}. Camera poses are denoted by matrices of the special Euclidean group $\mathbf{T}_i \in SE(3)$, which transform a 3D coordinate from the camera coordinate system to the world coordinate system. $\mathrm{\Pi}_{\mathbf{K}}$ and $\mathrm{\Pi}_{\mathbf{K}}^{\mathbf{-1}}$ denotes camera projection and back-projection functions, respectively. 

\subsubsection{Direct Image Alignment Formulation}
A point set $\mathcal{P}$ in the reference frame $I_{i}$ is observed in another frame $I_{j}$, and then the energy function of direct image alignment can be written as
\begin{equation}
\displaystyle
E_{ij} = \sum_{\mathbf{p} \in \mathcal{P}_{i}} \sum_{\mathbf{\tilde{p}} \in \mathcal{N}_{\mathbf{p}}} \omega_{\mathbf{\tilde{p}}} \left\lVert I_{j}[\mathbf{\tilde{p}^{\prime}}] - b_{j} - \frac{e^{a_{j}}}{e^{a_{i}}}\left(I_{i}[\mathbf{\tilde{p}}] - b_{i} \right) \right\rVert_{\gamma},
\label{eq:direct_align_energyfunc}
\end{equation}
where $\mathbf{p}$ is an image coordinate of a 3D point, $\mathcal{N}_{\mathbf{p}}$ is the 8-point pattern of $\mathbf{p}$, $||\cdot||_{\gamma}$ is a Huber norm, and $a_i, b_i, a_j, b_j$ model an affine brightness change for frame $i$ and $j$. The pattern point $\mathbf{\tilde{p}}$ is projected into $\mathbf{\tilde{p}^{\prime}}$ in $I_{j}$ calculated by
\begin{equation}
\displaystyle
\mathbf{\tilde{p}^{\prime}} = \mathrm{\Pi}_{\mathbf{K}} \left(\mathbf{T}_{ji} \mathrm{\Pi}_{\mathbf{K}}^{\mathbf{-1}}(\mathbf{\tilde{p}},d_{\mathbf{\tilde{p}}}) \right),
\end{equation}
where $d_{\mathbf{\tilde{p}}}$ is the inverse depth of $\mathbf{\tilde{p}}$, $\mathbf{T}_{ji}$ transforms a point from frame $i$ to frame $j$.
$\omega_{\mathbf{\tilde{p}}}$ is a down-weights high image gradients
\begin{equation}
\displaystyle
\omega_{\mathbf{\tilde{p}}} = \frac{c^2}{c^2 + \left\lVert \nabla    I_{i}(\mathbf{\tilde{p}}) \right\rVert_{2}^{2}},
\end{equation}
with a constant $c$.

\subsubsection{Tracking}
A semidense depth map is firstly estimated by static stereo matching for initialization of the visual odometry system. The inverse depth value of points is required by tracking the second frame by Equation (\ref{eq:direct_align_energyfunc}).

For tracking every new stereo frame, we project all the points inside the active window into the new stereo frame. We perform the optimization by minimizing the energy function of direct image alignment (\ref{eq:direct_align_energyfunc}) with Gauss-Newton. The pose of the new stereo frame is obtained by fixing the depth values.

\subsubsection{Frame Management}
After successfully tracking a new stereo frame, scene or illumination changing determines whether making a new keyframe or not. Mean squared optical flow is utilized to evaluate the scene changing between the new stereo frame and the last keyframe in the active window and quantized by the relative brightness.

If making a new keyframe, a sparse set of points is chosen from the current image, namely candidate points. For ensuring selected points with sufficient gradient across the images, we divided the image into small blocks. The size of small blocks is proportional to the size of the image. An adaptive threshold is calculated in each small block. If the gradient of a point is great enough to surpass the threshold, the point is selected.

If making no new keyframe, the non-keyframe is utilized to refine the inverse depth of candidate points. Static stereo matching acquires better tracking accuracy.

To control the growth of the size of the active window, old keyframes would be marginalized. The old active points that are hosted in the old keyframes and not observed by the two latest keyframes would be removed. After removing old keyframes and old points, candidate points become active points hosted in the new keyframe and observed in another keyframes, which prepares for creating new photometric energy function items for joint optimization.

\subsubsection{Joint Optimization}
Energy items of static stereo and temporal multi-view stereo are all considered in the joint optimization. The joint energy function can be written as 
\begin{equation}
\displaystyle
E = \sum_{i \in \mathcal{F}} \sum_{\mathbf{p} \in \mathcal{P}_{i}} \left( \sum_{j \in obs^t(\mathbf{p})} E_{ij}^{\mathbf{p}} + \lambda E_{is}^{\mathbf{p}} \right),
\label{eq:joint_opt_energyfunc}
\end{equation}
where $\mathcal{F}$ is the set of keyframes in the current window, $\mathcal{P}_i$ is the set of points in the frame $I_i$, $obs^t(\mathbf{p})$ are the observations of $\mathbf{p}$ from temporal multi-view stereo. $\lambda$ is a coupling factor between temporal multi-view stereo and static stereo.  
The energy function of temporal multi-view stereo $E_{ij}^{\mathbf{p}}$ can be described as
\begin{equation}
\displaystyle
E_{ij}^{\mathbf{p}} = \omega_{\mathbf{p}} \left\lVert I_{j}[\mathbf{p^{\prime}}(\mathbf{T}_i,\mathbf{T}_j,d,\mathbf{c})] - b_{j} - \frac{e^{a_{j}}}{e^{a_{i}}}\left(I_{i}[\mathbf{p}] - b_{i} \right) \right\rVert_{\gamma},
\end{equation}
and the energy function of static stereo $E_{is}^{\mathbf{p}}$ can be described as
\begin{equation}
\displaystyle
E_{is}^{\mathbf{p}} = \omega_{\mathbf{p}} \left\lVert I_{i}^{R}[\mathbf{p^{\prime}} ( \mathbf{T}_{ji}, d, \mathbf{c})] - b_{i}^{R} - \frac{e^{a_{j}^{R}}}{e^{a_{i}^{L}}}\left(I_{i}[\mathbf{p}] - b_{i}^{L} \right) \right\rVert_{\gamma},
\end{equation}
where $\mathbf{c}$ is the global camera intrinsics.
The final energy function is optimized by Gauss-Newton algorithm, which is further described in~\cite{engel_direct_2017, wang_stereo_2017}.

\subsubsection{Velocity Calculation}
After joint optimization of static stereo and temporal multi-view stereo in the active window, the current velocity is obtained from the poses of two latest stereo keyframes. A point is transformed from frame $I_i$ to frame $I_j$:
\begin{equation}
\displaystyle
\mathbf{T}_{ji} = \mathbf{T}_{j}^{-1}\mathbf{T}_i = 
\begin{bmatrix}
\mathbf{R}_{ji}   & t_{ji} \\
0       & 1 \\
\end{bmatrix}.
\end{equation}
\noindent
Rotational and translational velocity can be calculated from $\mathbf{R}_{ji}$ and $t_{ji}$, respectively. The rotational velocity $\omega$ can be written as
\begin{equation}
\displaystyle
\omega = \atantwo \left( \mathbf{R}_{ji}(2,0), \sqrt{\mathbf{R}_{ji}(2,1)^2 + \mathbf{R}_{ji}(2,2)^2 } \right) / \Delta t,
\end{equation}
and the translational velocity $v$ can be written as
\begin{equation}
\displaystyle
v = \sqrt{t_{ji}(0)^2 + t_{ji}(2)^2 } / \Delta t,
\end{equation}
where $\mathbf{R}_{ji}$ is 3-by-3 matrix, $v$ is a 3-by-1 vector.

\subsection{Calibration from Local View Cells}
Due to the accumulation of error during the process of path integration,  landmark calibrations are required for HD-by-Velocity cell network and grid-by-Velocity cell network. After visual calibrations, place cells help to build a coherent spatial encodes. For stereo camera images, local view templates are extracted from the left camera images to represent the current scenes. If a local view is never seen before, a new local view cell is created in the system, including the local view template, the HD units activities $m(\theta)$, and the three grid units activities $m(\vec{\theta})$ without velocity dimensions. If a local view is similar to one of the previous local view templates, the corresponding local view cell is activated. The activated local view cell injects energy into the HD-by-Velocity cell network and the three grid-by-Velocity cell networks. The HD units activities $m(\theta)$, and the three grid units activities $m(\vec{\theta})$ stored in the activated local view cell are scaled and expanded into velocity dimensions with the same values, and then, injected into the corresponding networks. The place cell activities are changed as the activity changes of three grid layers.

\subsection{Cognitive Map}
The mental map-like representation is learned in the entorhinal-hippocampal circuits, including HD cells, conjunctive grid cells, grid cells, and place cells. The spatial codes are read out to build cognitive maps by the experience map representation~\cite{ball_openratslam_2013}. A topological map is able to store the positions and their transitions according to direct sparse visual odometry. When the current place firing pattern is similar to the previous place firing pattern, a loop closure is adopted to optimize the topological map using a graph relaxation algorithm. 

\subsection{Implementation of the Cognitive Mapping Model}
\begin{figure}[!ht]
\begin{center}
\includegraphics[width=8cm]{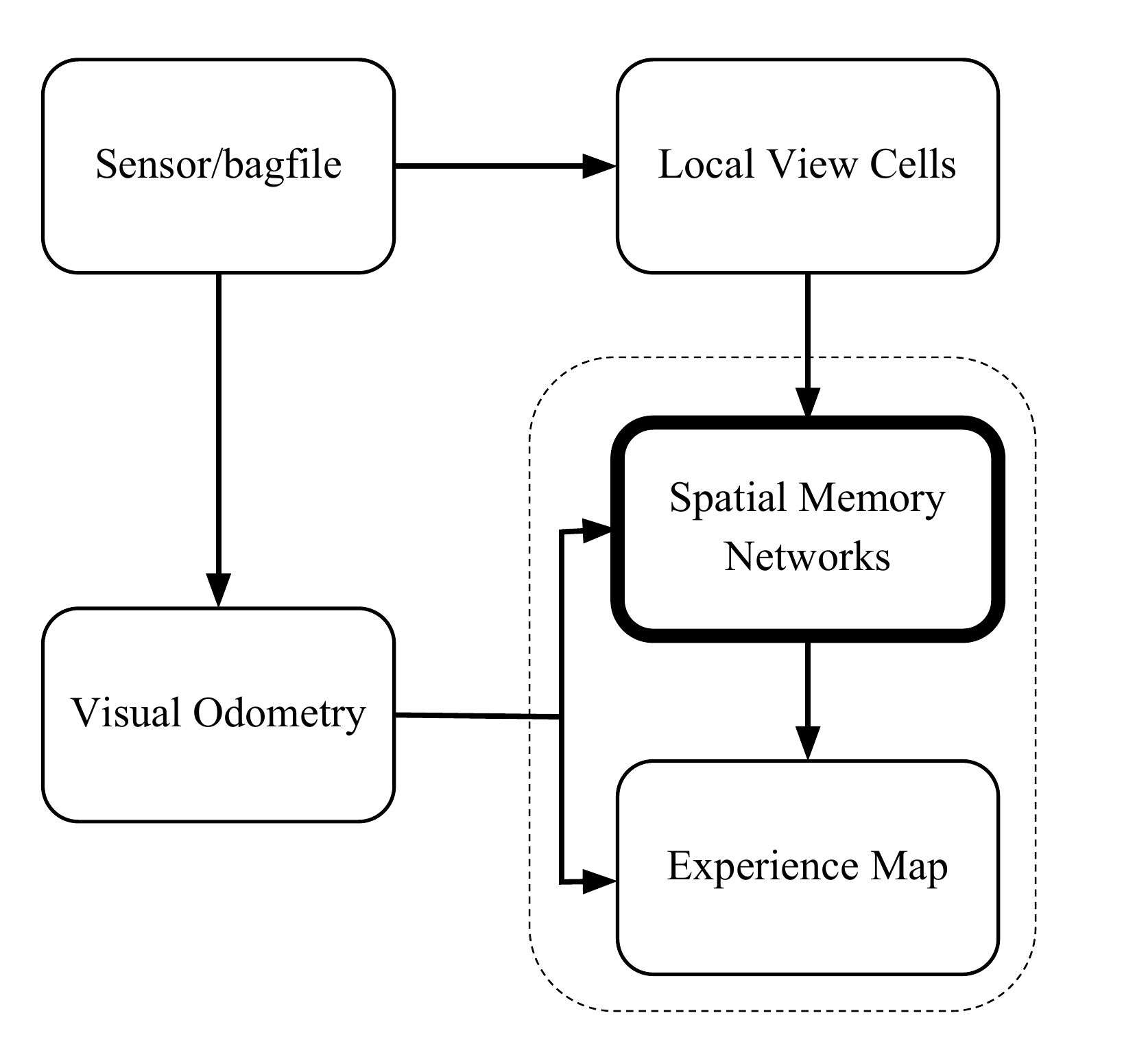} 
\end{center}
\caption{ The software architecture of the cognitive mapping system. The stereo images are provided by the sensor/bagfile node. Velocity is estimated by visual odometry node. The local view cell node determines whether the current view is novel or not. The spatial memory network node performs path integration and decision making to create links and vertices. The topological map is built by the experience map node.}\label{fig:node_structure}
\end{figure}

Our cognitive mapping system is implemented in Robot Operating System (ROS) Indigo on Ubuntu 14.04 LTS (Trusty) with C++ language. The software architecture of our cognitive mapping system is organized into five nodes (Figure~\ref{fig:node_structure}).

The visual odometry node real-time estimates the angular velocity and translational speed based on direction sparse method from a moving stereo camera. It receives images from ROS either from a camera or from stored data in a bagfile. 

The local view cell node determines whether the current image view is a novel or not. It provides calibration current to the networks in the spatial memory node. 

The spatial memory network node organically integrates HD cells, conjunction grid cells in the MEC layer III, V, and VI, grid cells in the MEC layer II, and place cells in the dentate gyrus together. This node receives two types of ROS messages as inputs: odometry and view templates. As shown in section~\ref{sec:v-input-hd} and~\ref{sec:v-input-grid}, the HD-by-Velocity cell network and grid-by-Velocity cell network integrate velocity information and visual information to form neural codes. The sparse place firing responses are generated in the dentate gyrus granule cells by the Winner-take-all mechanism with competitive Hebbian learning rules and controlling the sparsity of dentate gyrus activity. It distinctively encodes the pose of the robot. The spatial memory node also makes decisions about the creation of vertices and links in experience map, and sends ROS messages of graph operations to the experience map node.

The experience map node builds a coherent cognitive map from the neural codes of the place units. The key locations of the environment are represented as the vertices in a topological graph. A vertex stores the position estimated from the spatial memory network. A link maintains odometric transition information between vertices. On loop closure, a topological map relaxation method is used to find the minimum disagreement by optimizing the positions of vertices~\cite{duckett2002fast}. When the current position in the spatial memory network is far enough from the position of the previous vertex, a new vertex is created and a new edge is connected to the previous vertex.

\begin{figure*}[!ht]
\begin{center}
\includegraphics[width=12cm]{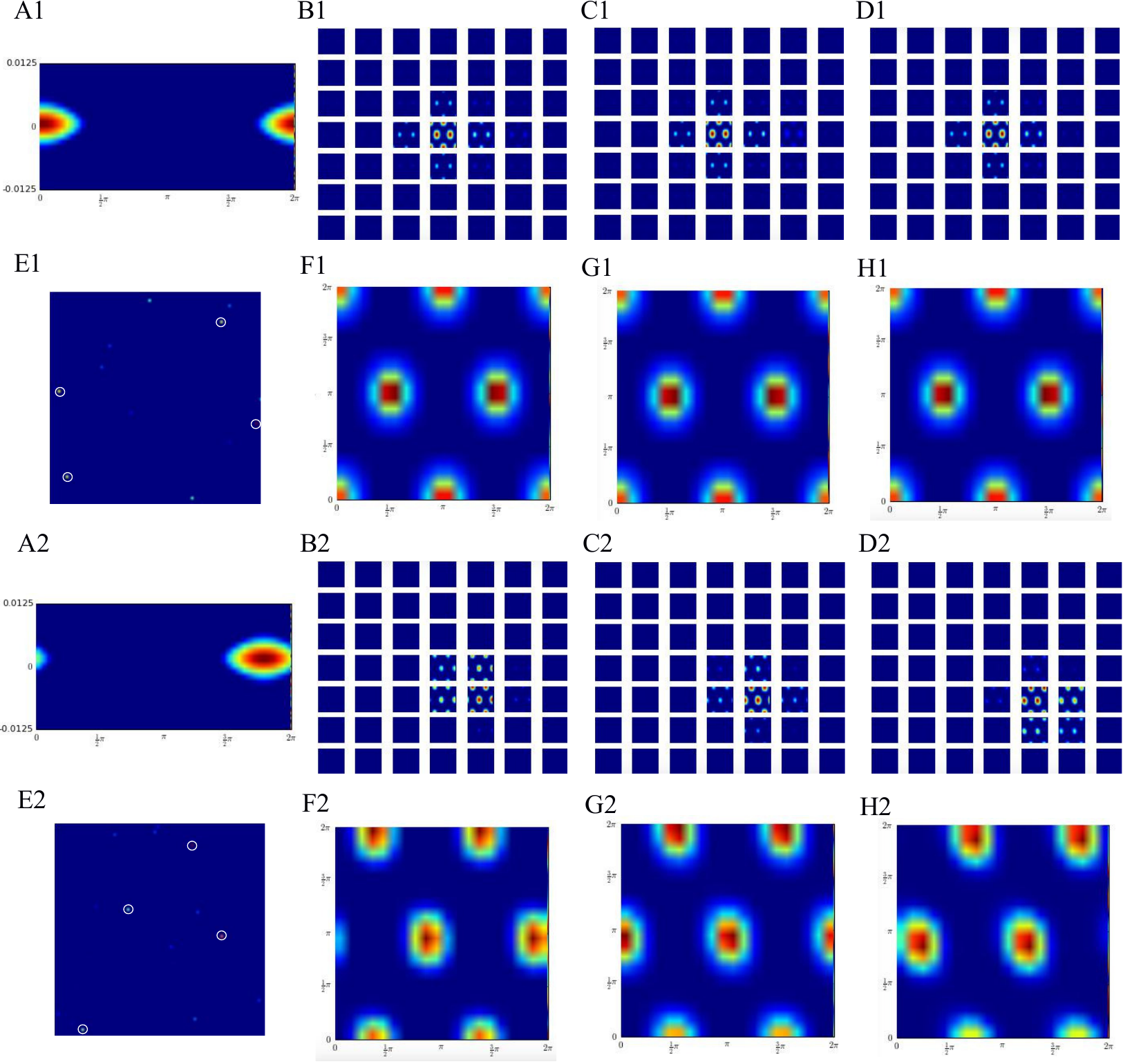} 
\end{center}
\caption{Neural representations in the cognitive mapping model. The activity of each unit from zero to maximal is color-coded from blue to red. Two groups of figures show activities of the units in the beginning of the experiments when the robot is static, and at a randomly selected time stamp when the robot is moving. (A) The population activity of the HD-by-velocity units. The direction $\theta$ shows on the horizontal axis. The velocity $\nu$ shows on the vertical axis. The center of the bump is at $(0,0)$ shown in A1. The center of the bump shown in A2 is at $(5.64, 0.0017)$. (B, C, D) The population activity of the grid-by-velocity units. In every conjunctive grid layer, the four-dimensional population activity is sliced in the $\nu_{x}$ and $\nu_{y}$ into 49 planes, $\nu_x$ increasing from left to right and $\nu_y$ decreasing from top to bottom. In each panel, the horizontal axis is the axis $\theta_{x}$ and the vertical axis is the axis $\theta_{y}$ with the same velocity label. Three same conjunctive grid patterns are shown in B1, C1, D1 in the beginning of the experiment. With velocity scale factor 1.4 between two consecutive grid modules, three different conjunctive grid patterns are shown in B2, C2, D2 in the middle of the experiment. (F, G, H) The population activity of the grid units is correspondingly summed from the population activity of the conjunctive grid units along velocity axes $\nu_{x}$ and $\nu_{y}$. (E) Sparse coding of place units. The units with stronger firing intensity in top 4 are labeled with white circles.}
\label{fig:neuralrepresentation}
\end{figure*}

\begin{figure*}[!ht]
\begin{center}
\includegraphics[width=16cm]{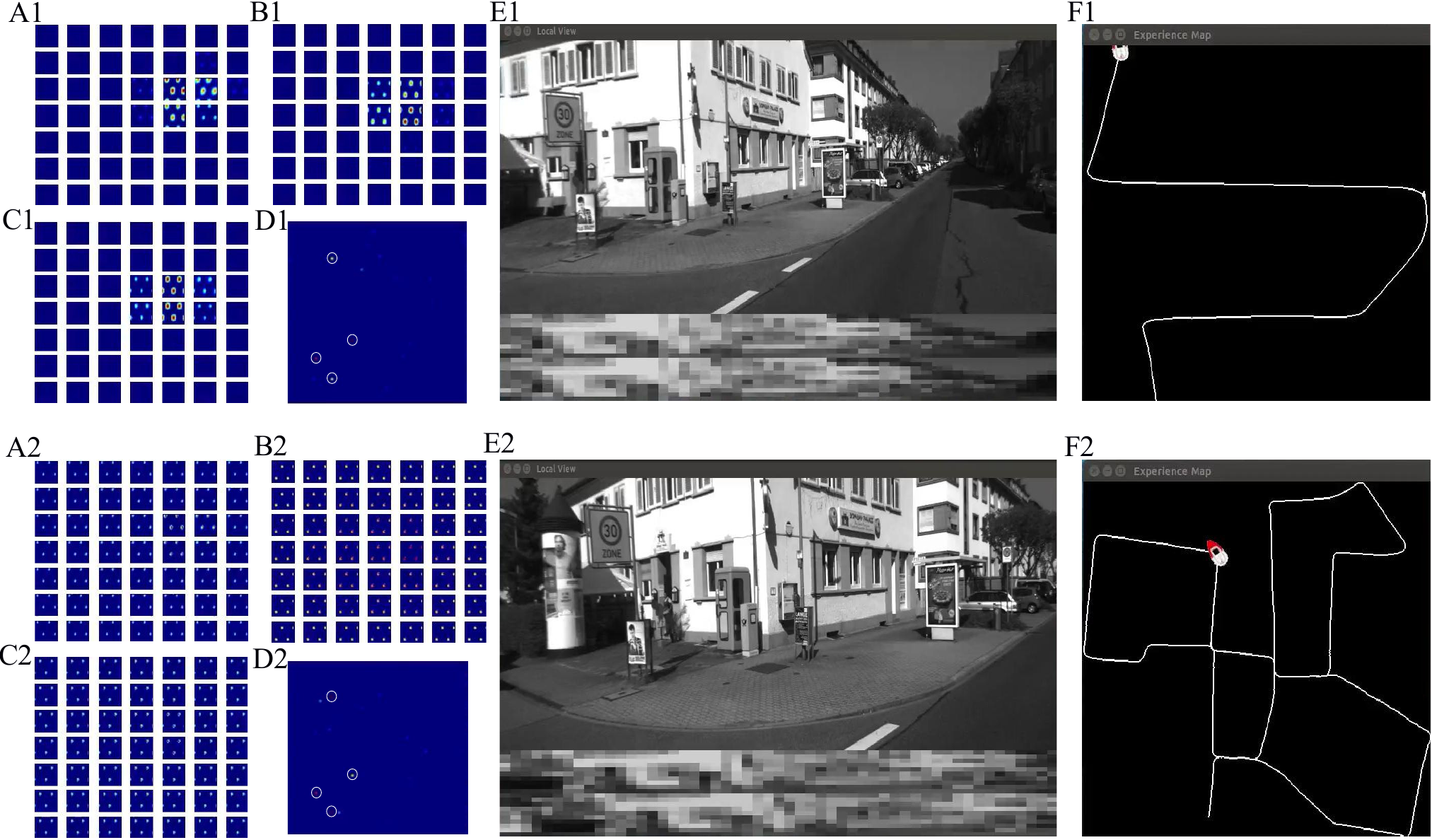} 
\end{center}
\caption{Stability of the place cell representation when the robot revisits a familiar location. There are two groups of figures. In each group, it includes three activity bumps of grid-by-velocity units (A, B, C), firing pattern of place units (D), the visual input (E), and experience map (F). (A, B, C) Three activity bumps of grid-by-velocity units are shown in A1, B1, and C1, when the robot visits the location first time. When the robot revisits the same location, neural network states of three grid-by-velocity units are calibrated by the visual feedbacks from local view cells. The local view cells inject currents into the three grid-by-velocity networks. After multiple continuous current injections, the grid patterns shown in A2, B2, and C2 become similar to the grid patterns shown in A1, B1, and C1 when the robot first visits the location. (D) The activity pattern of place units in D1 is similar to the pattern in D2, in terms of the top 4 units with stronger firing intensity. (E) The input image and matched template pair. E1 shows the same location as E2. (F) The current experience map. F1 and F2 shows the robot passes the corner for the first time and the second time, respectively. }
\label{fig:StabilityOfPlaceCellRepresentation}
\end{figure*}

Finally, we write python scripts to visualize the live state of our cognitive mapping system. In order not to trivially show the running windows of our system. The neural activity of HD-by-Velocity cells, grid-by-Velocity cells, grid cells and place cells can be found in Figure~\ref{fig:neuralrepresentation}. The image of the scene and the local view templates, as well as the current experience map,  can be found in Figure~\ref{fig:StabilityOfPlaceCellRepresentation}E and F, respectively. The mapping process shows all the live state of our cognitive mapping system in video S1 in Supplementary Materials.

\section{Results}
\label{results}

In this study, we resorted to robot navigation system for testing hypotheses about spatial cognition and validating our proposed cognitive mapping model. We modularly implemented our proposed cognitive mapping model based on ROS platform and demonstrated our model on the KITTI odometry benchmark dataset\cite{Geiger2012CVPR}. The KITTI odometry benchmark dataset is recorded by a stereo camera from a car with relatively high speed in urban and highway environments.
The stereo camera records images with resolution of $1241 \times 376$ pixels at 10 Hz. 
Our mapping system runs on a PC with 3.4 GHz six-core Intel i7 processor and 64 GB memory.
Video S1 in Supplementary Materials shows the mapping process of the KITTI odometry benchmark dataset sequence 00 by our implemented mapping system.

\subsection{Neural Representation}

Figure~\ref{fig:neuralrepresentation} shows two groups of activities of HD-by-velocity units (Figure~\ref{fig:neuralrepresentation}A), the grid-by-velocity units (Figure~\ref{fig:neuralrepresentation}B, C, and D), the grid units (Figure~\ref{fig:neuralrepresentation}F, G, and H), and the place units (Figure~\ref{fig:neuralrepresentation}E). 
Group 1 shows activities of all units in the beginning of the experiments, when the robot is stationary. At this moment, as angular velocity is zero, the bump of HD-by-velocity units is centered in the middle of the velocity dimension shown in Figure~\ref{fig:neuralrepresentation}A1. And since the inputs of translational velocities are also zero, three bumps of grid-by-velocity units are centered in the middle of velocity dimensions shown in Figure~\ref{fig:neuralrepresentation}B1, C1, and D1. When the robot is on the original position, three bumps of grid units express three same grid patterns in the initial state shown in Figure~\ref{fig:neuralrepresentation}F1, F1, and H1. With randomly initialized weights, after competitive Hebbian learning, the sparse firing pattern of place units is generated from grid units (in Figure~\ref{fig:neuralrepresentation}F1, F1, and H1) shown in Figure~\ref{fig:neuralrepresentation}E1. In order to easily show the results, the stronger firing units in top 4 are labeled with white circles.

And group 2 shows activities of all units at a randomly selected time stamp, when the robot is moving. The activity bump of HD-by-velocity is centered at (5.64, 0.0017) shown in Figure~\ref{fig:neuralrepresentation}A2, which means the robot is heading in $323.3^\circ$ and rotating at 0.17 rad/s. Considering the velocity scale factor is 1.4 between two consecutive grid-by-velocity modules in three modules, three activity bumps of grid-by-velocity moves with different velocity shown in Figure~\ref{fig:neuralrepresentation}B2, C2, and D2. With the same velocity input, three different grid patterns emerge to encode the same position in the physical environment shown in Figure~\ref{fig:neuralrepresentation}F2, F2, and H2. Competitive Hebbian learning with winner-take-all is adopted to generate a sparse firing pattern shown in Figure~\ref{fig:neuralrepresentation}E2. The two place firing patterns shown in Figure~\ref{fig:neuralrepresentation}E1 and E2 are nonoverlapping, which share no active place units, so that different locations can be easily distinguished.
It should be noted that the place units are listed randomly. For the convenience of viewing the firing pattern, the 6400 place units display in the form of 80-by-80 matrix.

Two different locations in the physical environment have two completely different sparse place firing pattern. When the robot revisits a familiar location, the place cell representation should be stable in different time stamp. Figure~\ref{fig:StabilityOfPlaceCellRepresentation} wants to show that when the robot revisits a location for the second time, the activity pattern of place units is sufficiently similar to the activity pattern of place units for the first time. From local view scenes (Figure~\ref{fig:StabilityOfPlaceCellRepresentation}E) and experience maps (Figure~\ref{fig:StabilityOfPlaceCellRepresentation}F), it can be easily confirmed that the robot revisited the same location. When the robot firstly comes to the location, three activity bumps of grid-by-velocity are shown in Figure~\ref{fig:StabilityOfPlaceCellRepresentation}A1, B1, and C1. After multiple continuous current injections to three grid-by-velocity networks from local view cells, the grid patterns in Figure~\ref{fig:StabilityOfPlaceCellRepresentation}A2, B2, and C2 are accordingly similar to the grid patterns in Figure~\ref{fig:StabilityOfPlaceCellRepresentation}A1, B1, and C1, without considering the velocity dimensions. 
For avoiding the influence of noise, place units with firing intensity in top 4 are just compared. 
The two sparse place firing patterns in Figure~\ref{fig:StabilityOfPlaceCellRepresentation}D1 and D2 present very high similarity at the same location in the physical environment, although the firing intensity of the place units is slightly different. In a word, the cognitive mapping model can encode the same location with the familiar place firing pattern, when the robot revisits a familiar location.

\subsection{Cognitive Map}

\begin{figure}[!ht]
\begin{center}
\includegraphics[width=8cm]{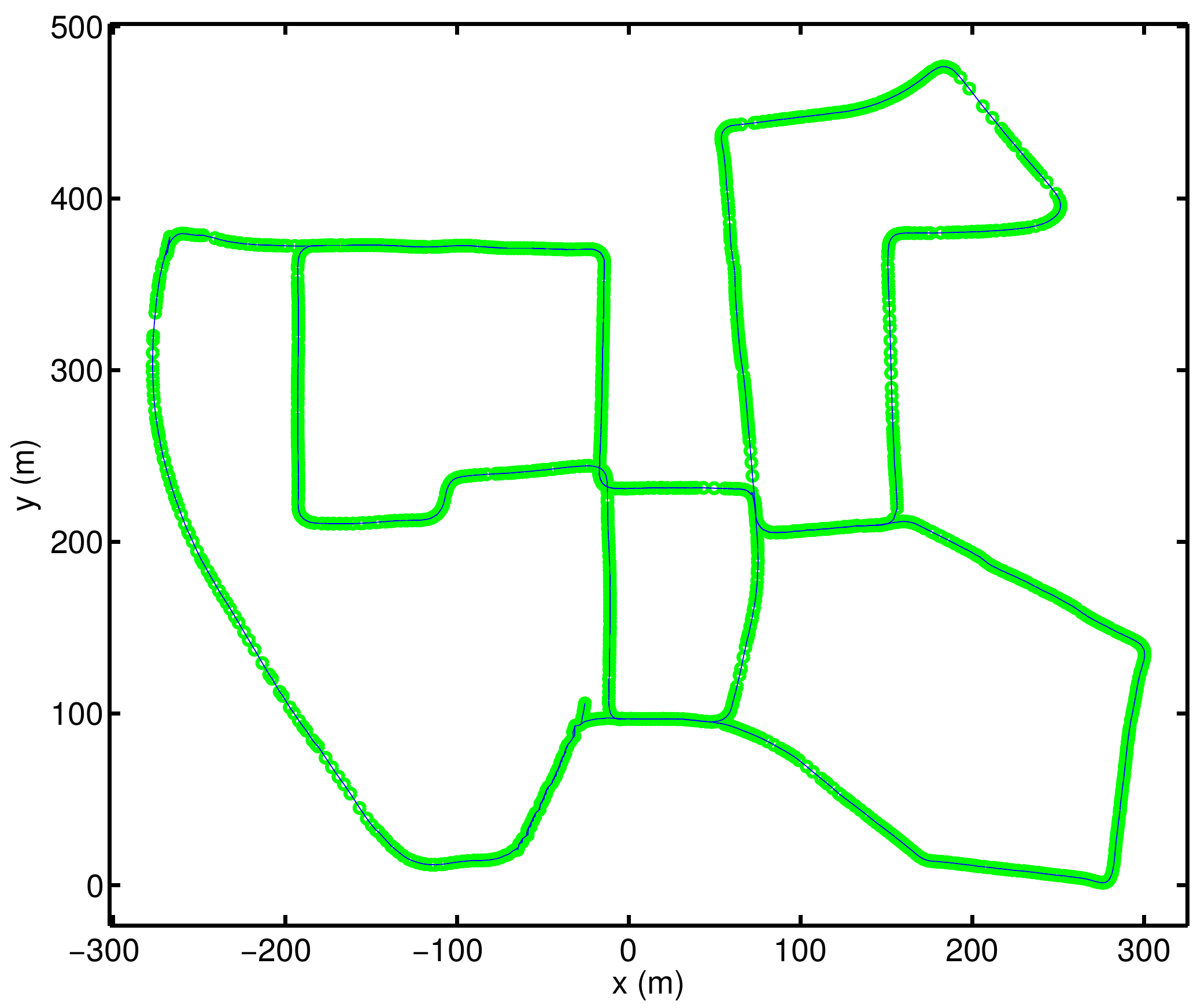} 
\end{center}
\caption{The cognitive map is a semi-metric topological map of the KITTI odometry benchmark dataset sequence 00 created by the mapping system. The topological vertices is presented by the
small green circles. The blue thin line describes links between connected vertices.}
\label{fig:kitti_cognitivemap}
\end{figure}

\begin{figure}[!ht]
\begin{center}
\includegraphics[width=8cm]{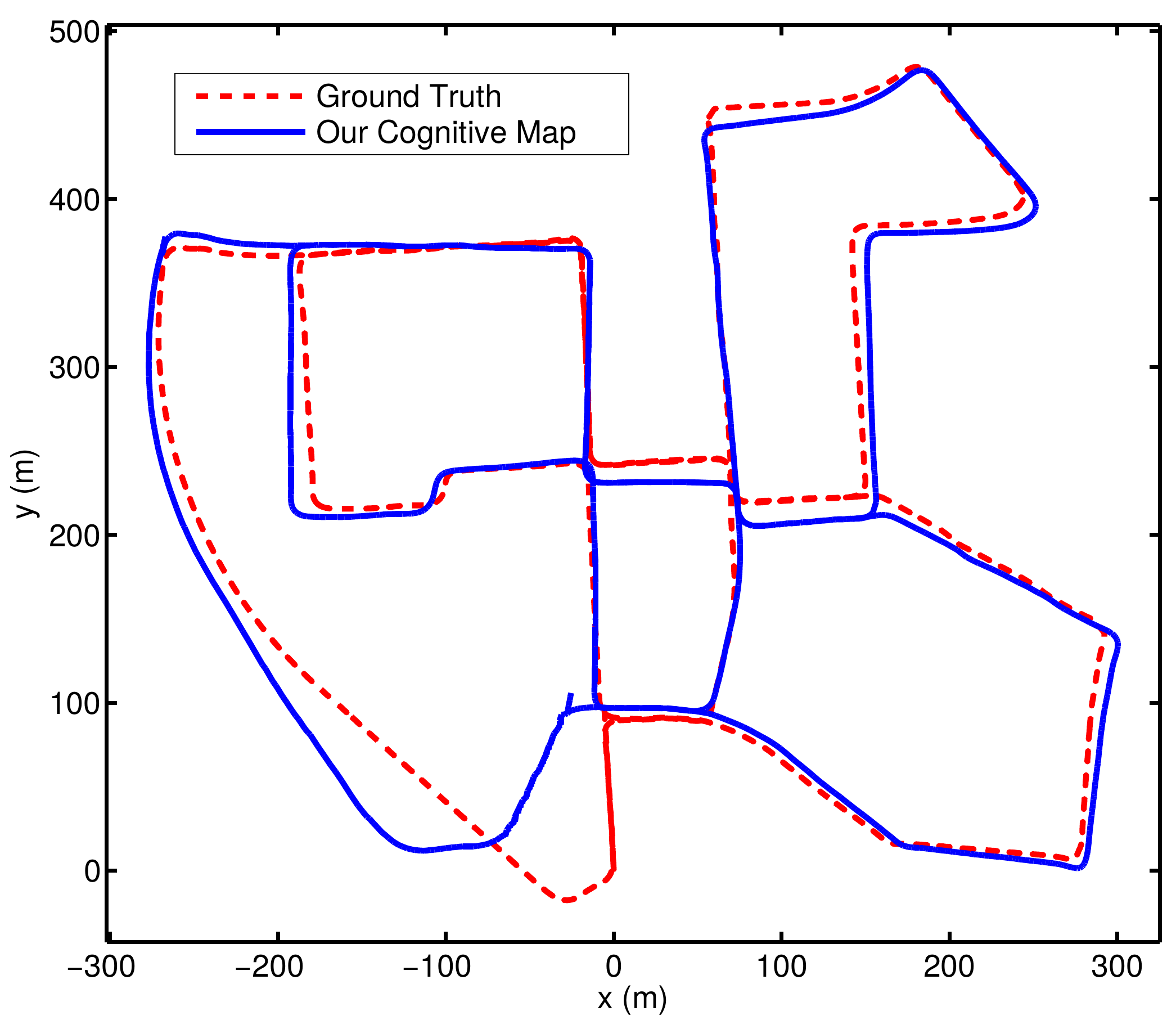} 
\end{center}
\caption{Our cognitive map (Blue) and ground truth (Red) of sequence 00 from the KITTI odometry benchmark dataset.}
\label{fig:kitti_cognitivemap_groundtruth}
\end{figure}

The cognitive map of the KITTI odometry benchmark dataset sequence 00 is generated by the implemented mapping system shown in Figure~\ref{fig:kitti_cognitivemap}. The vertices in the topological map are present by the thick green line, which represents the robot position in the explored environment. Two related vertices are connected by the link presented by the fine blue line.
As the link also contains the information about the physical distance, the experience map becomes a semi-metric topological map.

The cognitive map of the KITTI odometry benchmark dataset sequence 00 is qualitatively compared with the ground truth shown in Figure~\ref{fig:kitti_cognitivemap_groundtruth}.
The cognitive map captures the overall layout of the road network, including loop closures, intersections, corners, and curves intersections, which can be clearly seen using naked eyes. On the whole, the cognitive mapping system can build the cognitive map consistent with the ground truth of the environment.

\subsection{Firing Rate Maps}

\begin{figure}[!ht]
\begin{center}
\includegraphics[width=8cm]{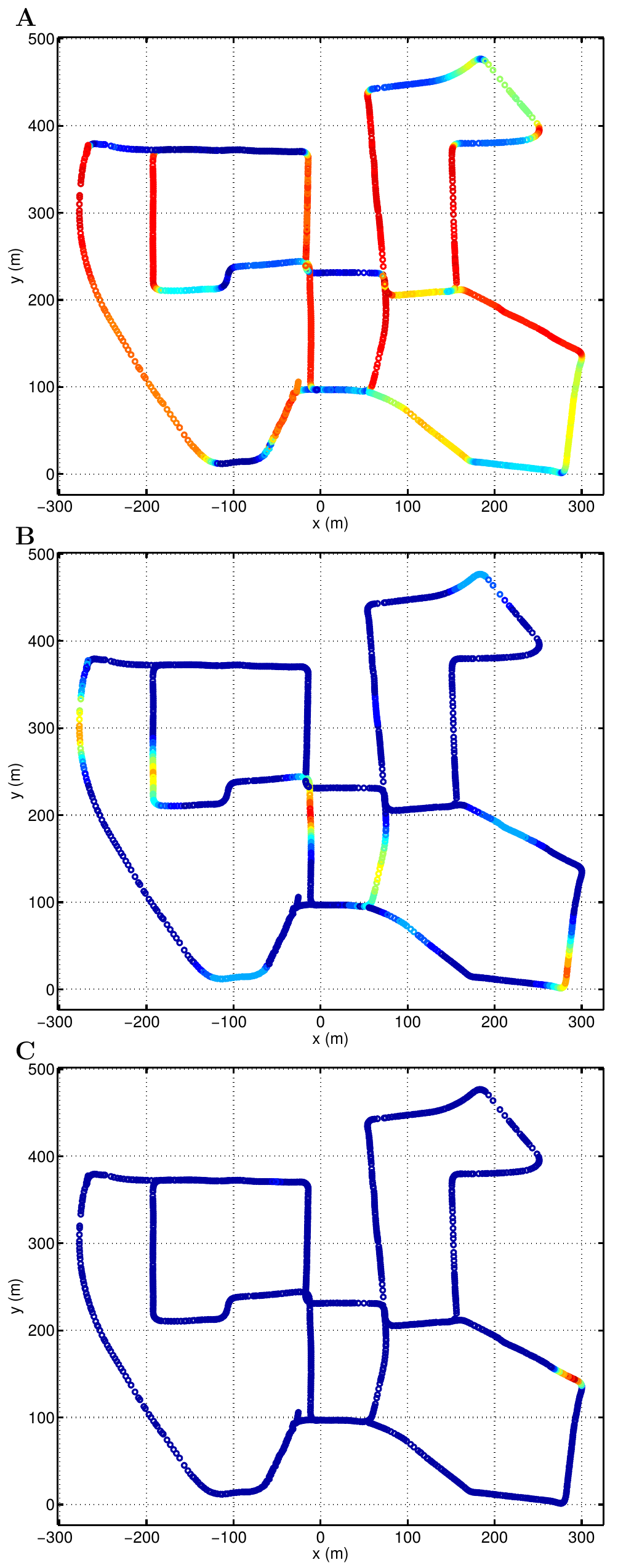} 
\end{center}
\caption{Firing rate maps of HD cells, grid cells, and place cells. In each panel, the firing rate is color-coded by the jet colormap from blue (zero firing rate) to red (high firing rate). (A) The firing rate map of the total activity of the HD units with two opposite preferred directions. (B) The firing rate map of a grid unit. (C) The firing rate map of a place unit.}
\label{fig:kitti_firingratemaps}
\end{figure}

The place units have already formed similar firing patterns to encode the same location in the physical environment when the robot revisits a location.
Here, in order to see the response of the single place unit, the activity of a place unit is shown on the top of the cognitive map. 
Figure~\ref{fig:kitti_firingratemaps}A shows the firing rate map of two HD units with two opposite preferred direction. The two HD units have strong firing rate only when the robot moves south to north or north to south on the parallel roads. Since the error accumulation during path integration process and the big width of the activity bump in the HD-by-velocity network, the two HD units also fire on the near road.
Due to the periodic boundary conditions on the torus attractor manifold, the grid unit fires at multiple locations in the explored environment shown in Figure~\ref{fig:kitti_firingratemaps}B. As the simple running trajectory of the robot is quite different from the repeated trajectory of the rat in the biological experiments, it would be unlikely to see the similar grid pattern in~\cite{hafting_microstructure_2005}. 

Then, the firing rate map of a place unit is shown in Figure~\ref{fig:kitti_firingratemaps}C. The neural representation of grid units and place units are completely different. Each grid unit is ordered and related to other grid units in the same grid layer by intrinsic attractor dynamics. Nevertheless, each place unit is arranged without order, and unrelated to other neighbor place units. The sparse firing patterns are formed by controlling the sparsity of the place units and competitive Hebbian learning with winner-take-all. The place unit locates at $100$ in the list of $6400$ place units. When the robot moves closer to the center of the place firing field presented in Figure~\ref{fig:kitti_firingratemaps}C, the firing rate alway gradually increases. Otherwise, when the robot moves farther to the center of the place firing field, the firing rate always slowly decreases. 
Since the grid unit fires at multiple locations in the environment, the ambiguity of neural representation exists in the grid cell network. Fortunately, the place unit only fires in one location, which can uniquely encode the location of the robot. Two different locations can be easily distinguished according to the firing patterns of place units.

\section{Discussion}
\label{discussion}

In this work, we develop an entorhinal-hippocampal model from grid cells to place cells for cognitive mapping. Three grid-by-velocity layers in the deep entorhinal cortex with different velocity gains are summed along velocity dimensions into three grid layers in the superficial entorhinal cortex and then projected into the dentate gyrus granule cells. The sparse place firing patterns are formed in the dentate gyrus granule cells by the Winner-take-all mechanism with competitive Hebbian learning rules and controlling the sparsity of dentate gyrus activity. Following the fruit fly olfactory algorithm~\cite{dasgupta_neural_2017}, three grid layers activities (400 grid cells in each layer) are expanded into the dentate gyrus granule cells (6400 granule cells) by a variant locality-sensitive hashing algorithm to sparsify the encoding of a location. 
We seek help from the robot system to validate our entorhinal-hippocampal model. We modularly implemented our entorhinal-hippocampal model on the vision-only robot mapping system based on ROS, and demonstrated on the KITTI odometry benchmark dataset, which is able to successfully build a coherent semi-metric topological map of the large-scale outdoor environment (see Video S1 in Supplementary Materials). 

In terms of the biological plausibility of the entorhinal-hippocampal model, in the rat hippocampus, there are 30,000 grid cells in the entorhinal cortex, and 1.2 million dentate gyrus granule cells in the hippocampus~\cite{dasgupta_neural_2017}. We basically followed the biological facts to design our model. If the size of the environment increases, more grid layers with diverse spacings, orientations, and phases, and dentate gyrus granule cells are needed to ensure that neural representations for two randomly selected locations share few, if any, active neurons to easily distinguish different locations. This function is called pattern separation in the dentate gyrus, which transforms relatively similar grid patterns into substantially different place firing patterns. 
From a computer science perspective, the entorhinal-hippocampal circuit can be thought of as a hashing function, whose inputs are three different grid patterns and whose output is a place firing pattern. The place firing pattern is a tag for the location (called hash). Although different place firing patterns are able to discriminate locations in the environment (Figure~\ref{fig:neuralrepresentation}), it also provides an ability to associate very similar locations with similar place firing patterns (Figure~\ref{fig:StabilityOfPlaceCellRepresentation}). When the robot comes back to the familiar scenes, a noisy place firing pattern is experienced. It is very likely to presume that the entorhinal-hippocampal circuit is locality-sensitive. The more similar two locations are, the more similar two place firing patterns are. We provide further evidence that this analogous algorithm from grid cells in the MEC to dentate gyrus granule cells in the hippocampus, namely, locality-sensitive hashing, may exist in the brain with different regions and species as a general mechanism of computation.

For the place firing patterns shown in Figure~\ref{fig:StabilityOfPlaceCellRepresentation},
two place firing patterns are just similar on the whole, not exactly the same pattern. This is because that multiple continuous current injections from local view cells make the sharpness of two grid patterns slightly different. And it further leads two place firing patterns to become slightly different.
As the error accumulation during the process of path integration, and local view cells anchor grids to local cues, the grid cells, and HD cells do not express firing patterns accurately shown in Figure~\ref{fig:kitti_firingratemaps}~\cite{derdikman2009fragmentation}. Without repeated exploration like the rat in the biological experiments, it is one of the reasons~\cite{hafting_microstructure_2005}.

The most relevant research is an entorhinal-hippocampal model using for cognitive map building~\cite{yuan_entorhinal-hippocampal_2015}. Population activity of place cells is created from multiple grid cell layers with different grid spacings in the neural space, not with different gains of the velocity input~\cite{mcnaughton_path_2006}. The population activity of place cells shows a single bump in the neural space. The specific region in the hippocampus is not determined for modeled the place cell. An amount of grid cell layers are used in their model to form place cell response with a simple bump in the neural space, whose number of grid cells is far more than the number of place cells. In our model, we used three grid layers with different gains of velocity inputs to form place firing patterns~\cite{mcnaughton_path_2006}, which specifically mimics the dentate gyrus granule cells in the hippocampus with place firing fields \cite{jonas_structure_2014}. The input patterns with a small number of grid cells are projected into very sparse firing patterns with a large number of place cells, which is realized by an algorithm analogous to locality-sensitive hashing~\cite{dasgupta_neural_2017}. 

Comparing with the fruit fly olfactory algorithm, odorant receptor neurons (ORNs) in the fly's nose are first converted into projection neurons (PNs) in the glomeruli, and then PNs are projected into Kenyon cells (KCs), connected by a sparse, binary random weight~\cite{dasgupta_neural_2017}. Whereas, in our work, the weight from grid cells to dentate gyrus granule cells is dense and learned by competitive Hebbian learning with the WTA mechanism. Since each location in the environment is encoded a place firing pattern determined only by a very small fraction of weight, the connection can also be considered to be sparse. 

Several limitations have remained in our study. 
First, although our cognitive mapping system has been demonstrated on the bench mark dataset, due to the large number of cells mimicked by our model, it is unlikely to run our system in real time.
Second, limited by computational resources, more grid layers with diverse spacings, orientations, and phases and granule cells in the dentate gyrus are not tested.  

For future works, we plan to quantitatively compare our topological maps with the ground truth of the environment, like metric maps. we will also further explore spatial representation of place cells in CA1 or CA3 regions in the hippocampus. 

\section{Conclusion}
\label{conclusion}
In a word, an entorhinal-hippocampal model from grid cells to place cells is proposed in this work to build the cognitive map based on the benchmark dataset. Three grid cell layers with different gains of velocity inputs are projected into the dentate gyrus granule cells to generate sparse place firing patterns for easily distinguishing different locations in the environment.
We view the entorhinal-hippocampal circuit as a variant locality-sensitive hashing function, like the fruit fly olfactory circuit. We further prove the algorithm may be a general principle of computation in different brain regions and species. Our model provides a better way to understand how spatial cognition works in the brain, and also develop models in analogous brain regions as a unified computing problem. Moreover, the model inspires a high-performance cognitive mapping system able to build a coherent semi-metric topological map on the benchmark dataset, and it also provides a possible alternative to build a more smarter and reliable robot brain from neuroscience.


\ifCLASSOPTIONcaptionsoff
  \newpage
\fi

\bibliographystyle{IEEEtran}
\bibliography{IEEEabrv,dg}





\vfill



\end{document}